\documentclass[aps,pra,amssymb,amsmath,eqsecnum,nofootinbib]{revtex4}
\usepackage{graphicx}

\newtheorem{definition}{Definition}[section]

\newtheorem{proposition}{Proposition}[section]

\newtheorem{conjecture}{Conjecture}[section]
\newtheorem{remark}{Remark}[section]
\newtheorem{axiom}{AXIOM}[section]

\newenvironment{hypothesis}{HP: \begin{center}} {\end{center}}
\newenvironment{thesis}{TH: \begin{center}} {\end{center}}
\newenvironment{proof}{\begin{center}PROOF: \end{center}} {$ \blacksquare $}

\begin{document}
\title{The multihistory approach to the time-travel paradoxes of General Relativity: mathematical analysis of a toy model}
\author{Gavriel Segre}
\begin{abstract}
With a mathematical eye to Matt Visser's multihistory approach to
the time-travel-paradoxes of General Relativity, a non
relativistic toy model is analyzed in order of characterizing the
conditions in which, in such a toy model, causation occurs.
\end{abstract}
\maketitle
\newpage
\tableofcontents
\newpage
\section{Causation versus chronology violations in General
Relativity} \label{sec:Causation versus chronology violations in
General Relativity}

 The only consistent definition of the concept of
causality is based on the Cauchy-Kowalewski theorem (see for
instance the section 1.7 of \cite{Courant-Hilbert-62b}) stating
that the initial value problem of a broad class of (partial)
differential equations has one and only one solution: in that case
one defines causation as the inter-relation existing between the
initial value and the later predicted value of the involved
quantity \footnote{Contrary to the typical philosophical attitude
of complicating simple things (see for instance \cite{Earman-86})
as to the definition of predictability we assume that the initial
condition may be known with exactness and we don't care about the
possible incomputability of the map expressing the final state
(the effect) as a function of the initial state (the cause).}.

\bigskip

 Let us remark that, in the general case, the
evolution parameter involved in the Cauchy problem doesn't
necessarily  coincide with the physical time.

\bigskip

Often it is is implicitly assumed the following:

\begin{conjecture} \label{con:conjecture of causation}
\end{conjecture}
\emph{Conjecture of Causation:}
\begin{center}
  The (partial) differential equations expressing the Laws of Nature have
  to admit, thanks to the Cauchy-Kowalewski theorem, a
  well-defined initial value problem.
\end{center}

\smallskip

Indeed   this is the situation to which we were used  since the
constraint involved in the conjecture \ref{con:conjecture of
causation} is satisfied by both Newtonian Mechanics  and Special
Relativity.

It is even satisfied by Quantum Mechanics where, beside the
nondeterministic reduction process, the evolution of a system non
subjected to measurements is ruled by the Schr\"{o}dinger equation
that of course admits a well-defined initial value problem.

\smallskip

One of the philosophical peculiarities of General Relativity
\cite{Hawking-Ellis-73}, \cite{Wald-84} is that, without assuming
ad hoc suppletive axioms in order to avoid such a situation, it
violates Conjecture\ref{con:conjecture of causation}: Einstein's
equation $ R_{ab} - \frac{1}{2} R g_{ab} \, = 8 \pi T_{ab} $
admits solutions $ ( M , g_{ab} ) $  that are not
globally-hyperbolic (i.e. that don't admit a Cauchy surface,
namely a closed achronal set S whose \emph{domain of dependence}
is the whole spacetime $ D(S) = M$):

in this case for every choice of a three-dimensional $C^{\infty} $
submanifold  S the predictability of the structure of spacetime
(and hence the causation) allowed from the initial value problem
on S is limited to the interior of the Cauchy-horizon of S:
\begin{equation}
    H_{Cauchy} ( S) \; := \; \partial D(S)
\end{equation}

The fact that in  such a situation the scalar wave equation may
have a well-defined Cauchy problem \cite{Friedman-04} is a very
little consolation.

An additional peculiarity of General Relativity  concerns
chronology violations:

given a time-orientable space-time $ ( M , g_{ab} ) $:
\begin{definition} \label{def:chronology violating set}
\end{definition}
\emph{chronology violating set of $ ( M , g_{ab} )$:}
\begin{equation}
    V_{chronology} ( M , g_{ab} ) \; := \; \cup_{p \in  M} I^{+} (p) \cap I^{-} (p)
\end{equation}

A curious feature of General Relativity is that there exist
solutions $( M , g_{ab} ) $ of Einstein's equation such that  $
V_{chronology} ( M , g_{ab} ) \; \neq \; \emptyset $.

Given such a solution and two points $ p_{1} , p_{2} \in M $:
\begin{definition} \label{def:chronological relation}
\end{definition}
\emph{chronological relation:}
\begin{equation}
    p_{1} \:  \preceq_{chronological} \: p_{2} \; := p_{2} \in I^{+}
    (p_{1}) \vee p_{1} = p_{2}
\end{equation}
Then:
\begin{proposition} \label{pr:properties of the chronological relation}
\end{proposition}
\begin{itemize}
    \item $   \preceq_{chronological} $ is a partial ordering relation over $ M
    -   V_{chronology} ( M , g_{ab} ) $
    \item  $   \preceq_{chronological} $  is a preordering
    over $  V_{chronology} ( M , g_{ab} )$ but
    is not a partial ordering
    \item
\begin{equation}
    \frac{ V_{chronology} ( M , g_{ab} )}{ \sim_{\preceq_{chronological}}} \;
    \; = \; \{ I^{+}(p) \cap I^{-}(p) \; p \in  V_{chronology} ( M , g_{ab}
    ) \}
\end{equation}
\end{itemize}
\begin{proof}
Let us observe, first of all, that
definition\ref{def:chronological relation} differs slightly from
the usual way through which the chronological relation
(traditionally denoted by $ << $) is defined in the literature
(see for instance the section 2.1 of \cite{Penrose-72},
\cite{Geroch-Horowitz-79} , the section3.2 "Causality Theory of
Space-times" of \cite{Beem-Ehrlich-Easley-96} and
\cite{Penrose-06}) that would correspond to our $ p_{1}
\prec_{chronological} p_{2} \; := \; p_{1} \preceq_{chronological}
p_{2} \, \wedge \, p_{1} \neq p_{2} $ \footnote{the chronological
relation may be in seen in some contexts as a conceptually more
fundamental mathematical structure than the metric (from which,
anyway, its definition obviously depends); an instance is given by
the definition of causal boundaries \cite{Harris-06} where other
kinds of non Hausdorffness, differing from those discussed in this
paper, occur.}; then:
\begin{equation}
 p \preceq_{chronological} p \; \; \forall p \in M
\end{equation}
 Demanding to the appendix \ref{sec:preorderings} for the involved
 notions concerning preorderings, the thesis follows by the fact
 that \cite{Hawking-Ellis-73} $ V_{chronology} ( M , g_{ab} ) $ is
 the disjoint union of sets of the form $ I^{+}(p) \cap I^{-}(p) $
 for $ p \in  V_{chronology} ( M , g_{ab} ) $ and that:
 \begin{equation}
  p_{1} \sim_{\preceq_{chronological}} p_{2} \; \; \forall p_{1} , p_{2} \in
  I^{+}(p) \cap I^{-}(p) , \forall p \in  V_{chronology} ( M , g_{ab} )
\end{equation}
\end{proof}

The situation delineated by proposition\ref{pr:properties of the
chronological relation} is often said to generate the so called
time travel paradoxes.

These paradoxes can be divided in two classes:
\begin{itemize}
    \item \emph{consistency paradoxes} involving the effects of
    the changes of the past (epitomized by the celebrated
    Grandfather Paradox in which a time-traveller goes back in the
    past and prevents the meeting of his grandfather and his
    grandmother)
    \item \emph{bootstrap paradoxes} involving the presence of
    loops in which the source of the production of some
    information disappears (as an example let us suppose that
    Einstein learnt  Relativity Theory from
    \cite{Hawking-Ellis-73},
    \cite{Wald-84} given to him by a time-traveller gone back to
    1904).
\end{itemize}

\begin{remark} \label{rem:time travel paradoxes and free will}
\end{remark}

As correctly stated in the section 6.4 "Causality conditions" of
\cite{Hawking-Ellis-73} these paradoxes occur only if one assumes
a simple notion of human free-will.

The existence of human free will is therein claimed to be a corner
stone of the Philosophy of Science underlying the Scientific
Method based on the assumption that one is free to perform any
experiment.

We don't agree with such a claim since, in our opinion, the
hypothetical assumption that we are determined to make the
experiments we perform is not incompatible with such a Philosophy
of Science.

Furthermore, at first sight, human free will would seem to be
incompatible with the determinism of General Relativity.

A deeper investigation about this claim  requires:
\begin{itemize}
    \item a precise
definition of both the concepts of determinism and free-will
    \item  the analysis whether, according to the assumed
    definitions of both the terms, determinism and free will
    are compatible (the position of Compatibilists) or not (the position of
    Incompatibilists) \footnote{We are using here the standard philosophical terminology
    \cite{Campbell-ORourke-Shier-04}; such a terminology introduces also the set of the Libertarians  defined as the subset of the Incompatibilists consisting in those believing in the existence of the free will while denying the existence of determinism. Apart from a general skeptisism about the
    whole philosophical approach to these themes, we regret the adoption of a term like libertarian, to which a positive ethical connotation is usually
    ascribed, to denote a conceptual position whose truth or falsity should be
neutrally decided analyzing the features of the Physical Theories
describing Nature.}
\end{itemize}

The determinism of a physical theory may be defined as the
condition that there exist dynamical equations (suitable (partial)
differential equations) governing the evolution of any closed
system.

Let us remark that, contrary to Laplace's classical definition
(see the celebrated second chapter "Concerning probability" of
\cite{Laplace-51}), such a definition of determinism doesn't
assume the Conjecture \ref{con:conjecture of causation}.

The simple notion of human free-will above mentioned consists in
the assumption that the true scientific theory describing human
mind is deterministic.

Obviously such a simple notion of free-will leads, by definition,
to the incompatibilist thesis.

There exist, anyway, more refined definitions of free-will that,
once assumed, make the compatibilistic thesis consistent.

\bigskip

The problem of the so called time-travel paradoxes has been faced
by the scientific community in different ways (see the fourth part
"Time Travel" of \cite{Visser-96} as well as \cite{Visser-03}):
\begin{enumerate}
    \item adding to General Relativity some ad hoc axiom
    precluding the physical possibility of causal loops (such as
    the strong form of Penrose's Cosmic Censorship Conjecture)
    \item appealing to consistency conditions (such as in Novikov's
    Consistency Conjecture) requiring that causal loops, though allowing causal influence on the past, don't allow
    alteration of the past
    \item arguing that the problem is removed at a quantum level (such as in Hawking's Chronology
    Protecting Conjecture stating that the  classical possibilities to implement time-travels are  destroyed by quantum effects)
    \item arguing that the so called time-travel paradoxes are only apparent and
    may be bypassed in a mathematical consistent way
\end{enumerate}

\bigskip

We have nor the knowledge neither the competence to take sides
about such a subtle issue.

\bigskip

In our opinion, anyway, of particular interest from a mathematical
viewpoint is the particular approach of the fourth kind according
to which the so called time-travel paradoxes are bypassed with the
removal of the assumption that space-time, as a topological space,
has to be Hausdorff (and hence allowing Universe's bifurcations in
multiple histories; see the section 19.1 "The radical rewrite
conjecture" of \cite{Visser-96} based on previous remarks by
Robert Geroch and Roger Penrose \cite{Penrose-79b} as well as by
Petr Hajicek \cite{Hajicek-71} \footnote{As it has happened many
times, the idea underlying such an approach has appeared first in
Science-Fiction's literature than in Science's literature: it is
for this reason that we suggest the lecture of the parts of
\cite{Nahin-01} concerning the logical analysis of time-travel
issues in the Science Fiction's literature.}) in that it allows an
interesting investigation about the topological structure of the
evolution parameter's space required in order that, eventually
under suitable consistency conditions, conjecture
\ref{con:conjecture of causation} holds.

\newpage
\section{Visser splittings}

Given a topological space (see appendix \ref{sec:Non Hausdorff
topological spaces})  $ ( X , \mathcal{T} ) $, a subset of its $
\Omega \subset X $ such that $
\partial \Omega \; \neq \;
\emptyset $ and a natural number $ n \in \mathbb{N} : n \geq 2 $:
\begin{definition}
\end{definition}
\emph{Visser n-splitting of X through $ \Omega $:}
\begin{equation}
    Split(X, \Omega , n ) \; := \; [ X - \bar{\Omega} ] \cup [
    \cup_{i=1}^{n} \bar{\Omega}_{i} ]
\end{equation}
where $ \Omega_{1} , \cdots , \Omega_{n} $ are n disjoint copies
of $ \Omega $.

\begin{definition} \label{def:natural topology on a Visser splitting}
\end{definition}
\emph{natural topology of $  Split(X, \Omega , n ) $:}
\begin{center}
 the topology $ \tilde{\mathcal{T}} (X, \Omega , n ) $ having the
 following basis: any open set in $ [ X - \bar{\Omega} ] \cup
\bar{ \Omega}_{i} , i=1, \cdots, n $ is an open set of  $
\tilde{\mathcal{T}} (X, \Omega , n )
 $.
\end{center}

From here and beyond we will assume that any Visser splitting is
endowed with its natural topology.

It is important to observe that:
\begin{proposition}
\end{proposition}
\emph{non Hausdorffness of Visser splittings:}

\begin{center}
 the topological space $ (  Split(X, \Omega , n ) , \tilde{\mathcal{T}} (X, \Omega , n
 ) ) $ is not Hausdorff
\end{center}
\begin{proof}
   Given $ i,j=1, \cdots , n \: : \: i \neq j
   $, let us consider two points $  x_{i} \in
    \partial \Omega_{i} , x_{j} \in
    \partial \Omega_{j}  $ such that $ x_{i} $ and $ x_{j} $ are copies of the same element $ x \in \partial
    \Omega $; the definition \ref{def:natural topology on a Visser
   splitting} implies that $  x_{i} \curlyvee x_{j} $.
\end{proof}

\smallskip

Let us now suppose that $ \preceq $ is a partial ordering over X
(see appendix \ref{sec:preorderings}). Given $ x , y  \in Split(X,
\Omega , n )$:
\begin{definition} \label{def:preordering on a Visser splitting}
\end{definition}
\begin{equation}
    x \widetilde{\preceq} y \; := \; x \preceq y
\end{equation}
Then:
\begin{proposition} \label{pr:preordering on a Visser splitting}
\end{proposition}
\begin{center}
\begin{itemize}
    \item  $ \widetilde{\preceq} $ is a preordering over $  Split(X,
\Omega , n ) $
    \item $ \widetilde{\preceq} $ is not a partial ordering over $  Split(X,
\Omega , n ) $
\end{itemize}
\end{center}
\begin{proof}
 The reflexive and the transitive property of  $ \widetilde{\preceq}
 $ may be immediately inferred by the corresponding properties of
 $ \preceq $.

$ \widetilde{\preceq} $ doesn't satisfy, anyway, the antisymmetric
property since given $ x \in \Omega $ and considered the
corresponding copies $ x_{i} \in \Omega_{i} \; i=1 , \cdots , n $:
\begin{equation}
   ( x_{i} \tilde{\preceq} x_{j} \: \wedge \: x_{j} \widetilde{\preceq}
    x_{i} \; \nRightarrow \; x_{i} = x_{j} ) \; \;  i, j =
    1 , \cdots , n \, : \, i \neq j
\end{equation}
\end{proof}

\begin{remark}
\end{remark}

Gordon Mc Cabe has recently \cite{Mc-Cabe-05} introduced the
following alternative to Visser splitting:
\begin{definition} \label{def:Mc Cabe splitting}
\end{definition}
\emph{Mc Cabe n-splitting of X through $ \Omega $:}
\begin{equation}
    Split_{Mc \; Cabe} ( X , \Omega , n ) \; := \;  \frac{Split ( X , \Omega , n )}{\sim_{Mc \; Cabe} }
\end{equation}
where $ \sim_{Mc \; Cabe} $ is the equivalence relation over $
Split ( X , \Omega , n ) $ defined by:
\begin{equation}
    x_{i} \sim_{Mc \; Cabe} x_{j} \; \Leftrightarrow \; x_{i} \in
    \partial \Omega_{i} , x_{j} \in
    \partial \Omega_{j}  \wedge \text{ $ x_{i} $ and $ x_{j} $ are copies of the same element $ x \in \partial \Omega
    $} \; \; i,j=1, \cdots , n \: : \: i \neq j
\end{equation}

\begin{definition} \label{def:natural topology on a Mc Cabe splitting}
\end{definition}
\emph{natural topology of $  Split_{Mc \; Cabe}(X, \Omega , n )
$:}
\begin{equation}
    \tilde{\mathcal{T}}_{Mc \; Cabe} (X, \Omega , n ) \; := \; \frac{\tilde{\mathcal{T}} (X, \Omega , n ) }{ \sim_{Mc \; Cabe}}
\end{equation}

Then:
\begin{proposition}
\end{proposition}
\begin{equation}
   \sim_{Mc \; Cabe} \text { is Hausdorff }
\end{equation}
\begin{proof}
 Since:
 \begin{equation}
    x \curlyvee y \; \Rightarrow \; x \sim_{Mc \; Cabe} y \; \;
    \forall x,y \in Split ( X , \Omega , n )
\end{equation}
the thesis follows
\end{proof}

\smallskip

Actually Mc Cabe's formalism is nothing but an application of the
strategy indicated by Geroch in the exercise 177 of the $ 27^{th}
$ section "Continuous Mappings" of \cite{Geroch-85} :
\begin{center}
 "make equivalent as few points as necessary to get an Hausdorff
 quotient space"
\end{center}
For this reason we won't adopt Mc Cabe splittings in this paper.

\newpage
\section{Visser splittings of time and Initial Value Problems: a toy model}

We will consider for simplicity from here and beyond a particular
toy model consisting of a classical non-relativistic dynamical
system having as configuration space the real line $ ( \mathbb{R}
, T_{natural} ( \mathbb{R} )  ) $ and whose evolution parameter t
(that in this particular case will hence coincide with the
physical absolute time) takes values on some Visser splitting of
the real line  $ ( Split(\mathbb{R}, \Omega , n ) ,
\tilde{\mathcal{T}} (\mathbb{R}, \Omega , n )  )$.

The usual linear ordering $ \leq $ over $ \mathbb{R} $ induces, by
proposition \ref{pr:preordering on a Visser splitting}, the
preordering $ \tilde{\leq} $ over $ Split(\mathbb{R}, \Omega , n )
$ that, in according to the explained underlying physical
interpretation, we will denote by $ \preceq_{chronological} $.

Given a map $ x : ( Split(\mathbb{R}, \Omega , n ) ,
\tilde{\mathcal{T}} (\mathbb{R}, \Omega , n )  ) \mapsto (
\mathbb{R} , + , \cdot ) $, according to the analysis performed in
the appendix \ref{sec:Differential Calculus on non Hausdorff
topological spaces} it is well defined the concept of time
derivative of $ x( t) $ that we will denote by $ \dot{x} (t) $.

Given a map $ f \in C^{\infty} (  \mathbb{R} ) $, a point $ t_{in}
\in  Split(\mathbb{R}, \Omega , n )$ and a real number $ x_{in}
\in \mathbb{R} $ we will investigate under which conditions on $
\Omega $ the initial value problem:
\begin{equation} \label{eq:evolution equation}
 \dot{x} \; = \; f(x)
\end{equation}
\begin{equation} \label{eq:initial condition}
    x( t_{in} ) \; = \; x_{in}
\end{equation}
is well-posed, i.e. it admits one and only one solution $ x(t) : (
Split(\mathbb{R}, \Omega , n ) , \tilde{\mathcal{T}} (\mathbb{R},
\Omega , n )  ) \mapsto ( \mathbb{R} , + , \cdot ) $.

\smallskip

Let us start considering the simplest case $ \Omega := \{ 0 \} , n
:= 2 $.

Let us start from the case $ t_{in} \notin \{ 0_{1} , 0_{2} \} $.

Called $ \tilde{x}(t) : \mathbb{R} \mapsto \mathbb{R} $ the
solution of the Cauchy problem  eq. \ref{eq:evolution equation},
eq. \ref{eq:initial condition} for ordinary time $ t \in
\mathbb{R} $:

\begin{proposition}
\end{proposition}

\begin{hypothesis}
\end{hypothesis}
\begin{equation}
   t_{in} \in Split( \mathbb{R} , \{ 0 \} , 2 ) - \{ 0_{1} , 0_{2} \}
\end{equation}
\begin{thesis}
\end{thesis}

\begin{center}
 The Cauchy problem   eq.  \ref{eq:evolution equation}, eq.
\ref{eq:initial condition} for $ t \in Split ( \mathbb{R} , \{ 0
\} , 2 \} $ is  well-defined, its unique solution being the map $
\tilde{\tilde{x}} (t) : Split ( \mathbb{R} , \{ 0 \} , 2 ) \mapsto
\mathbb{R} $:
\end{center}
\begin{equation}
  \tilde{\tilde{x}} (t) \; := \; \tilde{x} (t) \; \forall t \in
  \mathbb{R} - \{ 0 \}
\end{equation}
\begin{equation}
   \tilde{\tilde{x}} ( 0_{1} ) \; := \; \tilde{x} ( 0)
\end{equation}
\begin{equation}
      \tilde{\tilde{x}} ( 0_{2} ) \; := \; \tilde{x} (0)
\end{equation}
\begin{proof}
The thesis immediately follows by the definition
\ref{def:derivative} and the structure of the topology
 $ \tilde{\mathcal{T}} (\mathbb{R}, \{ 0 \} , 2 )$.
\end{proof}

\bigskip

Let us now suppose that $ t_{in} :=  0_{1}  $ and let $
\tilde{x}(t) : \mathbb{R} \mapsto \mathbb{R} $ be the solution of
the Cauchy problem:
\begin{equation}
 \dot{x} \; = \; f(x)
\end{equation}
\begin{equation}
    x( 0 ) \; = \; x_{in}
\end{equation}
Then:
\begin{proposition}
\end{proposition}

\begin{hypothesis}
\end{hypothesis}
\begin{equation}
    t_{in} :=  0_{1}
\end{equation}

\begin{thesis}
\end{thesis}

\begin{center}
 The Cauchy problem   eq.  \ref{eq:evolution equation}, eq.
\ref{eq:initial condition} for $ t \in Split ( \mathbb{R} , \{ 0
\} , 2 \} $ is  well-defined, its unique solution being the map $
\tilde{\tilde{x}} (t) : Split ( \mathbb{R} , \{ 0 \} , 2 ) \mapsto
\mathbb{R} $:
\end{center}
\begin{equation}
  \tilde{\tilde{x}} (t) \; := \; \tilde{x} (t) \; \forall t \in
  \mathbb{R} - \{ 0 \}
\end{equation}
\begin{equation}
   \tilde{\tilde{x}} ( 0_{1} ) \; := \; \tilde{x} ( 0)
\end{equation}
\begin{equation}
      \tilde{\tilde{x}} ( 0_{2} ) \; := \; \tilde{x} (0)
\end{equation}
\begin{proof}

It follows from the definition \ref{def:derivative} and the
structure of the topology
 $ \tilde{\mathcal{T}} (\mathbb{R}, \{ 0 \} , 2 )$.
\end{proof}

\bigskip

The generalization to $ n \in \mathbb{N} : n > 2 $ is
straightforward.

\bigskip

Let us now consider the case in which $ \Omega := [ 0 , + \infty )
, n := 2 $.

If $ t_{in} \in ( - \infty , 0) $ let $ \tilde{x}(t) : \mathbb{R}
\mapsto \mathbb{R} $ be the solution of the Cauchy problem  eq.
\ref{eq:evolution equation}, eq. \ref{eq:initial condition} for
ordinary time $ t \in \mathbb{R} $.

Then:
\begin{proposition} \label{pr:initial value before splittings}
\end{proposition}

\begin{hypothesis}
\end{hypothesis}
\begin{equation}
     t_{in} \in ( - \infty , 0)
\end{equation}
\begin{thesis}
\end{thesis}

\begin{center}
 The Cauchy problem   eq.  \ref{eq:evolution equation}, eq.
\ref{eq:initial condition} for $ t \in Split ( \mathbb{R} , [ 0 ,
+ \infty ) , 2 ) $ is  well-defined, its unique solution being the
map $ \tilde{\tilde{x}} (t) : Split ( \mathbb{R} , [ 0 , + \infty
) , 2 ) \mapsto \mathbb{R} $:
\end{center}
\begin{equation}
  \tilde{\tilde{x}} (t) \; := \; \tilde{x} (t) \; \forall t \in (
  - \infty , 0 )
\end{equation}
\begin{equation}
   \tilde{\tilde{x}} ( t_{1} ) \; := \; \tilde{x} ( t = t_{1} ) \; \;
   \forall t_{1} \in [ 0_{1} , + \infty_{1} )
\end{equation}
\begin{equation}
   \tilde{\tilde{x}} ( t_{2} ) \; := \; \tilde{x} ( t = t_{2} ) \; \;
   \forall t_{2} \in [ 0_{2} , + \infty_{2} )
\end{equation}

\begin{proof}
The thesis immediately follows by the definition
\ref{def:derivative} and the structure of the topology
 $ \tilde{\mathcal{T}} (\mathbb{R}, \{ 0 \} , 2 )$.
\end{proof}

\smallskip

Given $ a,b \in \mathbb{R} $ let us consider the generalized
Cauchy problem:
\begin{equation} \label{eq:repeated evolution equation}
    \dot{x} = f(x)
\end{equation}
\begin{equation}  \label{eq:first initial condition}
    x( 0_{1} ) = a
\end{equation}
\begin{equation} \label{eq:second initial condition}
    x( 0_{2} ) = b
\end{equation}

Given an arbitrary $ r \in \mathbb{R} $ let $ \tilde{x}(t)_{r} :
\mathbb{R} \mapsto \mathbb{R} $ be the solution of the Cauchy
problem:
\begin{equation}
    \dot{x} = f(x)
\end{equation}
\begin{equation}
    x(0) = r
\end{equation}
for ordinary time $ t \in \mathbb{R} $.

Then:
\begin{proposition} \label{pr:initial value in splitting points}
\end{proposition}
The generalized Cauchy problem eq.\ref{eq:repeated evolution
equation}, eq.\ref{eq:first initial condition}, eq. \ref{eq:second
initial condition} for  $ t \in Split ( \mathbb{R} , [ 0 , +
\infty ) , 2 ) $ is well-defined if and only if $a=b $;  in this
case the unique solution is the map $ \tilde{\tilde{x}} (t) :
Split ( \mathbb{R} , [ 0 , + \infty ) , 2 ) \mapsto \mathbb{R} $
such that:
\begin{equation}
    \tilde{\tilde{x}} (t) := \tilde{x}(t)_{a=b} \; \; \forall t \in
    ( - \infty , 0 )
\end{equation}
\begin{equation}
  \tilde{\tilde{x}} (t_{1}) :=  \tilde{x}(t=t_{1})_{a=b} \; \;
  \forall t_{1} \in [ 0_{1} , + \infty_{1} )
\end{equation}
\begin{equation}
  \tilde{\tilde{x}} (t_{2}) :=  \tilde{x}(t=t_{2})_{a=b} \; \;
  \forall t_{2} \in [ 0_{2} , + \infty_{2} )
\end{equation}
\begin{proof}
 The retrodiction to $ ( - \infty , 0)  $ is possible if and only
 if:
 \begin{equation}
    \tilde{x}(t)_{a} = \tilde{x}(t)_{b} \; \; \forall t \in  ( - \infty , 0)
\end{equation}
and   hence if and only if $a = b$.
\end{proof}

\smallskip

Let us now suppose that $ t_{in} \in [ 0_{1} , + \infty_{1} ) $
and let $ \tilde{x} (t) : \mathbb{R} \mapsto  \mathbb{R} $ be the
solution of the Cauchy problem  eq. \ref{eq:evolution equation},
eq. \ref{eq:initial condition}  for ordinary time $ t \in
\mathbb{R} $. Then:

\begin{proposition} \label{pr:initial value after splittings}
\end{proposition}

\begin{hypothesis}
\end{hypothesis}
\begin{equation}
    t_{in} \in [ 0_{1} , + \infty_{1} )
\end{equation}
\begin{thesis}
\end{thesis}
The Cauchy problem eq. \ref{eq:evolution equation}, eq.
\ref{eq:initial condition} for $ t \in  Split ( \mathbb{R} , [ 0 ,
+ \infty ) , 2 ) $ is well-defined, its unique solution being the
map $ \tilde{\tilde{x}} (t) : Split ( \mathbb{R} , [ 0 , + \infty
) , 2 ) \rightarrow \mathbb{R} $:
\begin{equation}
    \tilde{\tilde{x}} ( t) \; := \; \tilde{x}(t) \; \; \forall t \in
    ( - \infty , 0 )
\end{equation}
\begin{equation}
      \tilde{\tilde{x}} ( t_{1}) \; := \; \tilde{x} ( t = t_{1} )
      \; \; \forall t_{1} \in [ 0_{1} , + \infty_{1} )
\end{equation}
\begin{equation}
      \tilde{\tilde{x}} ( t_{2}) \; := \; \tilde{x} ( t = t_{2} )
      \; \; \forall t_{2} \in [ 0_{2} , + \infty_{2} )
\end{equation}
\begin{proof}
 The solution on the first branch $ [ 0_{1} , + \infty_{1} ) $
 allows the retrodiction on $ ( - \infty , 0 ) $ from which the
 prediction on $ [ 0_{2} , + \infty_{2} ) $ can be derived.
\end{proof}
\newpage
\section{The toy model with a tree-like topological structure of time}
The situation discussed in the previous section may be easily
generalized to the case of multiple Visser splitting of time:

given $ n \in \mathbb{N}_{+} $, $ t_{1} , \cdots , t_{n} \in
\mathbb{R} $ such that $ t_{1} < t_{2} < \cdots < t_{n} $, n
natural numbers greater or equal than two $ b_{1} , b_{2} , \cdots
, b_{n} \in \mathbb{N} : b_{i} \geq 2 \; \; i=1, \cdots , n$ and
$n-1$ natural numbers $ i_{1} \in \{ 1 , 2, \cdots , b_{1} \} ,
\cdots , i_{n-1} \in \{ 1 , 2, \cdots , b_{n} \} $:
\begin{definition} \label{def:temporal tree}
\end{definition}
\emph{temporal tree:}
\begin{multline}
    tree(t_{1} , b_{1} ; t_{2,(i_{1})} , b_{2} ;  \cdots ; t_{n,(i_{n-1})} , b_{n} ) \; :=
     \\
   Split ( \cdots Split( Split ( \mathbb{R} , [ t_{1} , + \infty )  , b_{1} ) , [
   t_{2,(i_{1})} , + \infty_{(i_{1})})
   , b_{2} ) \cdots , [
   t_{n,(i_{n-1})} , + \infty_{(i_{n-1})}) , b_{n} )
\end{multline}

\smallskip

\begin{remark}
\end{remark}
It important not to make confusion between the two different kind
of labels: those denoting the time ordering and those denoting the
different branches; to avoid confusion we will denote this second
kind of label enclosing it between brackets.

So the first time splitting occurs at $ t_{1} $ of which we will
have $ b_{1} $ copies that we will denote by $ t_{1 ,(1)} , \cdots
, t_{1 ,(b_{1})} $ and so on.

\smallskip

\begin{remark}
\end{remark}
Let us remark that the iterated adoption of the definition
\ref{def:natural topology on a Visser splitting} induces a
topology on $  tree(t_{1} , b_{1} ; t_{2,(i_{1})} , b_{2} ; \cdots
; t_{n,(i_{n-1})} , b_{n} )$ that we will denote by $
\mathcal{T}(t_{1} , b_{1} ; t_{2,(i_{1})} , b_{2} ;  \cdots ;
t_{n,(i_{n-1})} , b_{n} ) $.

\smallskip

\begin{remark} \label{rem:difference between a temporal tree and its graph}
\end{remark}
The name of definition \ref{def:temporal tree} may be someway
misleading and has to be managed carefully.

Its origin arises observing that introduced the following:
\begin{definition} \label{def:graph of a temporal tree}
\end{definition}
\emph{graph of $  tree(t_{1} , b_{1} ; t_{2,(i_{1})} , b_{2} ;
\cdots ; t_{n,(i_{n-1})} , b_{n} )$}:

\begin{center}
 the oriented graph \cite{Bollobas-98} $ graph[  tree(t_{1} , b_{1} ; t_{2,(i_{1})} , b_{2} ;  \cdots ; t_{n,(i_{n-1})} , b_{n} )] $ obtained by $  tree(t_{1} , b_{1} ; t_{2,(i_{1})} , b_{2} ;  \cdots ; t_{n,(i_{n-1})} , b_{n} ) $  replacing each element  $ [ t_{i-1,(j)} \cdots , t_{i,(j)} )
    \cup \cup_{k=1}^{b_{i}} [ t_{i,(k)} , t_{i+1,(k)} ) $ with a vertex $
    t_{i} $ having one entering edge corresponding to the interval
    $ [ t_{i-1,(j)} \cdots , t_{i,(j)} ) $ and having $ b_{i} $ exiting
    edges corresponding to the intervals $ [ t_{i,(1)} ,
    t_{i+1,(1)}) , \cdots , [ t_{i,(b_{i})} ,
    t_{i+1,(b_{i})}) $ (where we have denoted by $ t_{(0)} \in ( - \infty , t_{1})  $ the
    time parameter before the first splitting)
\end{center}

$ graph  [tree(t_{1} , b_{1} ; t_{2,(i_{1})} , b_{2} ; \cdots ;
t_{n,(i_{n-1})} , b_{n} )] $ is indeed a tree.

It is important, anyway, to remark that $ graph  [  tree(t_{1} ,
b_{1} ; t_{2,(i_{1})} , b_{2} ; \cdots ; t_{n,(i_{n-1})} , b_{n}
)] $ is only a diagrammatic way of representing the different
mathematical object $ tree(t_{1} , b_{1} ; t_{2,(i_{1})} , b_{2} ;
\cdots ; t_{n,(i_{n-1})} , b_{n} ) $ that is not a graph and hence
is not, in particular, a tree.

\bigskip

Given a map $ f \in C^{\infty} (  \mathbb{R} ) $, an initial time
$ t_{in}  \in tree(t_{1} , b_{1} ; \cdots ; t_{n} , b_{n} )$ and a
real number $ x_{in} \in \mathbb{R} $:
\begin{proposition} \label{pr:ordinary Cauchy problem for temporal trees}
\end{proposition}

The Cauchy problem:
\begin{equation} \label{eq:evolution equation again}
 \dot{x} \; = \; f(x)
\end{equation}
\begin{equation} \label{eq:initial condition again}
    x( t_{in} ) \; = \; x_{in}
\end{equation}
is well-posed i.e. it admits one and only one solution

 $ \tilde{x}(t) : (
 tree(t_{1} , b_{1} ; t_{2,(i_{1})} , b_{2} ; \cdots
; t_{n,(i_{n-1})} , b_{n} ), \mathcal{T}(t_{1} , b_{1} ;
t_{2,(i_{1})} , b_{2} ; \cdots ; t_{n,(i_{n-1})} , b_{n} ))
\mapsto ( \mathbb{R} , + , \cdot ) $

\begin{proof}
 The thesis immediately follows by multiple application of
 Proposition \ref{pr:initial value before splittings} and
 Proposition  \ref{pr:initial value after splittings}.
\end{proof}

\bigskip

Given $ i \in \{ 1 , \cdots , n  \} $ and $ a_{1} , \cdots
a_{b_{i}} \in \mathbb{R} $ let us now consider the generalized
Cauchy problem:
\begin{equation} \label{eq:again repeated evolution equation}
    \dot{x} = f(x)
\end{equation}
\begin{equation}  \label{eq:multiple initial condition}
    x( t_{i,(1)} ) = a_{1} , \cdots,  x( t_{i,(b_{i})} ) = a_{b_{i}}
\end{equation}
Then:
\begin{proposition}
\end{proposition}
\begin{center}
 The generalized Cauchy problem of eq. \ref{eq:again repeated evolution
 equation}, eq. \ref{eq:multiple initial condition} is well-posed
 if and only if $ a_{1} = \cdots = a_{b_{i}} $
\end{center}

\begin{proof}

The thesis immediately follows by multiple application of
 Proposition \ref{pr:initial value in splitting points}.
\end{proof}
\newpage
\section{The toy model with a less trivial topological structure of time}
Given the Visser splitting $ Split ( \mathbb{R} , ( - \infty , 0 ]
, 2 ) $, a  map $ f \in C^{\infty} ( \mathbb{R} ) $, $ t_{in} \in
( - \infty , 0 ] $ and $ a_{1}, a_{2} \in \mathbb{R} $ let us
consider the generalized Cauchy problem:
\begin{equation} \label{eq:once more evolution equation}
    \dot{x} = f(x)
\end{equation}
\begin{equation}  \label{eq:once more first initial condition}
    x( t_{in,(1)} ) = a_{1}
\end{equation}
\begin{equation}  \label{eq:once more second initial condition}
    x( t_{in,(2)} ) = a_{2}
\end{equation}
As usual we will say that the Cauchy problem of eq. \ref{eq:once
more evolution equation}, eq. \ref{eq:once more first initial
condition} and eq. \ref{eq:once more second initial condition} is
well-posed if and only if it has one and only one solution $
\tilde{x} : ( Split ( \mathbb{R} , ( - \infty , 0 ] , 2 ) ,
\tilde{\mathcal{T}} ( \mathbb{R} , ( - \infty , 0 ] , 2 ) )
\mapsto ( \mathbb{R}, + , \cdot)$.

Then:
\begin{proposition}
\end{proposition}
\begin{center}
 The Cauchy problem of eq. \ref{eq:once
more evolution equation}, eq. \ref{eq:once more first initial
condition} and eq. \ref{eq:once more second initial condition} is
well-posed if and only if $ a_{1} = a_{2} $.
\end{center}
\begin{proof}
The condition $ a_{1} = a_{2} $ is necessary and sufficient for
the merging of the solutions in the two branches $ (- \infty_{1},
0_{1} ] $ and  $ (- \infty_{2}, 0_{2} ] $  in $ ( 0 , + \infty )$.
\end{proof}

\bigskip

We can now combine the Visser splittings of the form  $ Split (
\mathbb{R} , [ t_{i} , + \infty ) , n ) $ and $ Split ( \mathbb{R}
, [ - \infty ,  t_{i}) , n ) $ to obtain  more intricate
topological structures of time:

given $ n \in \mathbb{N}_{+} $, $ t_{1} , \cdots , t_{n} \in
\mathbb{R} $ such that $ t_{1} < t_{2} < \cdots < t_{n} $, n
natural numbers greater or equal than two $ b_{1} , b_{2} , \cdots
, b_{n} \in \mathbb{N} : b_{i} \geq 2 \; \; i=1, \cdots , n$,
$n-1$ natural numbers $ i_{1} \in \{ 1 , 2, \cdots , b_{1} \} ,
\cdots ,  i_{n-1} \in \{ 1 , 2, \cdots , b_{n} \} $ and n boolean
variables $ k_{1}, \cdots, k_{n} \in \{ 0 , 1 \}$:
\begin{definition} \label{def:chronologically preordered temporal structure}
\end{definition}
\emph{chronologically preordered temporal structure:}
\begin{multline}
    structure(t_{1} , b_{1} , k_{1} ; t_{2,(i_{1})} , b_{2} , k_{2} ;  \cdots ; t_{n,(i_{n-1})} , b_{n} , k_{n} ) \; :=
     \\
   Split ( \cdots Split( Split ( \mathbb{R} , \Omega(t_{1})  , b_{1} ) , \Omega_{t_{2,(i_{1})}} t_{2,(i_{1})} ,  b_{2} ) \cdots , [
   \Omega_{t_{n,(i_{n-1})}} , b_{n} ))
\end{multline}
where:
\begin{equation}
    \Omega(t_{1}) := \left\{%
\begin{array}{ll}
    [ t_{1} , + \infty ), & \hbox{if $ k_{1} = 0$;} \\
    ( - \infty , t_{1} ], & \hbox{if $ k_{1} = 1$} \\
\end{array}%
\right.
\end{equation}
\begin{equation}
    \Omega(t_{2}) := \left\{%
\begin{array}{ll}
     [ t_{2,(i_{1})} , + \infty_{(i_{1})}), & \hbox{if $ k_{2} = 0$;} \\
     ( - \infty_{(i_{1})}) , t_{2,(i_{1})} ] , & \hbox{if $ k_{2} = 1$} \\
\end{array}%
\right.
\end{equation}
\begin{equation*}
 \vdots
\end{equation*}
\begin{equation}
    \Omega(t_{n}) := \left\{%
\begin{array}{ll}
     [ t_{n,(i_{n-1})} , + \infty_{(i_{n-1})}), & \hbox{if $ k_{n} = 0$;} \\
     ( - \infty_{(i_{n-1})} , t_{n,(i_{n-1})} ]  , & \hbox{if $ k_{n} = 1$} \\
\end{array}%
\right.
\end{equation}

The elements of $ structure(t_{1} , b_{1} , k_{1} ; t_{2,(i_{1})}
, b_{2} , k_{2} ;  \cdots ; t_{n,(i_{n-1})} , b_{n} , k_{n} ) $
with $ k_{i} = 0 $ will be called \emph{time-divisions} while the
elements of $ structure(t_{1} , b_{1} , k_{1} ; t_{2,(i_{1})} ,
b_{2} , k_{2} ;  \cdots ; t_{n,(i_{n-1})} , b_{n} , k_{n} ) $ with
$ k_{i} = 1 $ will be called \emph{time-stickings}.

\bigskip

\begin{remark}
\end{remark}
The terminology adopted in the definition \ref{def:chronologically
preordered temporal structure} is owed to the fact that the
multiple application of definition \ref{def:preordering on a
Visser splitting} induces on  $ structure(t_{1} , b_{1} , k_{1} ;
t_{2,(i_{1})} , b_{2} , k_{2} ;  \cdots ; t_{n,(i_{n-1})} , b_{n}
, k_{n} ) $ the chronological relation $ \preceq_{chronological} $
that, by the multiple application of the proposition
\ref{pr:properties of the chronological relation}, is a
preordering.

\bigskip

\begin{remark}
\end{remark}
Let us remark that the iterated adoption of the definition
\ref{def:natural topology on a Visser splitting} induces a
topology on $ structure(t_{1} , b_{1} , k_{1} ; t_{2,(i_{1})} ,
b_{2} , k_{2} ;  \cdots ; t_{n,(i_{n-1})} , b_{n} , k_{n} )$ that
we will denote by $ \mathcal{T}( t_{1} , b_{1} , k_{1} ;
t_{2,(i_{1})} , b_{2} , k_{2} ;  \cdots ; t_{n,(i_{n-1})} , b_{n}
, k_{n} )$.

\bigskip

\begin{definition} \label{def:graph of a temporal structure}
\end{definition}
\emph{graph of $ structure(t_{1} , b_{1} , k_{1} ; t_{2,(i_{1})} ,
b_{2} , k_{2} ;  \cdots ; t_{n,(i_{n-1})} , b_{n} , k_{n} ) $:  }

\begin{center}
 the oriented graph $ graph[ structure(t_{1} , b_{1} , k_{1} ; t_{2,(i_{1})} ,
b_{2} , k_{2} ;  \cdots ; t_{n,(i_{n-1})} , b_{n} , k_{n} ) ] $
obtained by $ structure(t_{1} , b_{1} , k_{1} ; t_{2,(i_{1})} ,
b_{2} , k_{2} ;  \cdots ; t_{n,(i_{n-1})} , b_{n} , k_{n} ) $
\begin{itemize}
    \item replacing each \emph{time-division}  $ \langle t_{i-1,(j)} \cdots , t_{i,(j)} )
    \cup \cup_{k=1}^{b_{i}} [ t_{i,(k)} , t_{i+1,(k)} \rangle  $ with a vertex $
    t_{i} $ having one entering edge corresponding to the interval
    $ \langle t_{i-1,(j)} \cdots , t_{i,(j)} ) $ and having $ b_{i} $ exiting
    edges corresponding to the intervals $ [ t_{i,(1)} ,
    t_{i+1,(1)} \rangle , \cdots , [ t_{i,(b_{i})} ,
    t_{i+1,(b_{i})} \rangle $
    \item  replacing each \emph{time-sticking} $  \cup_{k=1}^{b_{i}} \langle t_{i-1,(k)} , t_{i,(k)} ]
     \cup ( t_{i,(j)} , t_{i+1,j} \rangle $ with a vertex $
    t_{i} $ having $ b_{i} $ entering
    edges corresponding to the intervals $ \langle t_{i-1,(1)} , t_{i,(1)}
    ] , \cdots ,  \langle t_{i-1,(b_{i})} , t_{i,(b_{i})}
    ] $ and having one exiting edge  corresponding to the interval
    $ ( t_{i,(j)} , t_{i+1,j} \rangle $
\end{itemize}
where we have denoted by "$ \langle $" a parenthesis that can be a
"$ ( $" or a "$ [ $"  while we have denoted by "$ \rangle $" a
parenthesis that can be a "$ ) $" or a "$ ] $".
\end{center}

\bigskip

\begin{remark} \label{rem:difference between a temporal structure and its graph}
\end{remark}
Such as definition \ref{def:chronologically preordered temporal
structure} is a generalization of definition \ref{def:temporal
tree} so definition \ref{def:graph of a temporal structure}  is a
generalization of definition \ref{def:graph of a temporal tree}.

As we have already done in the remark \ref{rem:difference between
a temporal tree and its graph} as to temporal trees and their
graphs we remark here that $ structure(t_{1} , b_{1} , k_{1} ;
t_{2,(i_{1})} , b_{2} , k_{2} ;  \cdots ; t_{n,(i_{n-1})} , b_{n}
, k_{n} ) $ and its graph $ graph [ structure(t_{1} , b_{1} ,
k_{1} ; t_{2,(i_{1})} , b_{2} , k_{2} ;  \cdots ; t_{n,(i_{n-1})}
, b_{n} , k_{n} ) ] $ are two distinct mathematical notions.

\bigskip

\begin{remark}
\end{remark}
Let us observe that, taking into account the orientation of edges,
$ graph[ structure(t_{1} , b_{1} , k_{1} ; t_{2,(i_{1})} , b_{2} ,
k_{2} ;  \cdots ; t_{n,(i_{n-1})} , b_{n} , k_{n} ) ] $ doesn't
contain loops.

\bigskip

Given a map $ f \in C^{\infty} (  \mathbb{R} )$, an initial time $
t_{in} \in structure(t_{1} , b_{1} , k_{1} ; t_{2,(i_{1})} , b_{2}
, k_{2} ; \cdots ; t_{n,(i_{n-1})} , b_{n} , k_{n} ) $ and a real
number $ x_{in} \in \mathbb{R} $:
\begin{proposition}
\end{proposition}

The Cauchy problem:
\begin{equation} \label{eq:evolution equation more again}
 \dot{x} \; = \; f(x)
\end{equation}
\begin{equation} \label{eq:initial condition more again}
    x( t_{in} ) \; = \; x_{in}
\end{equation}
is well-posed i.e. it admits one and only one solution

 $ \tilde{x}(t) : (
 structure(t_{1} , b_{1} , k_{1} ;
t_{2,(i_{1})} , b_{2} , k_{2} ;  \cdots ; t_{n,(i_{n-1})} , b_{n}
, k_{n}  , b_{n} ), \mathcal{T}(t_{1} , b_{1} ; t_{2,(i_{1})} ,
b_{2} ; \cdots ; t_{n,(i_{n-1})} , b_{n} )) \mapsto ( \mathbb{R} ,
+ , \cdot ) $

\begin{proof}
  The thesis immediately follows by Proposition \ref{pr:ordinary Cauchy problem for
  temporal trees}.
\end{proof}
\newpage
\section{The introduction of chronological circularities in the toy model}

In this section we will consider topologies of time giving rise to
causal circularities.

Let us start from the simplest case:

given the topological space $ ( \mathbb{R} , T_{natural} (
\mathbb{R} ) ) $ and two points $ t_{1} , t_{2} \in \mathbb{R} :
t_{1} < t_{2} $:

\begin{definition} \label{def:identification of two points on the line}
\end{definition}
\emph{identification of $ t_{1} $ and $ t_{2}$ on $ \mathbb{R} $:}
\begin{equation}
    Identification( \mathbb{R} ; t_{1} , t_{2} ) \; := \; \frac{( \mathbb{R} , T_{natural} (
\mathbb{R} ))}{\sim_{t_{1},t_{2}}}
\end{equation}
where $ \sim_{t_{1},t_{2}} $ is the equivalence relation over $
\mathbb{R} $:
\begin{equation}
    a \sim_{t_{1},t_{2}} b \; := \; \exists n \in \mathbb{Z} :
    b = a + n ( t_{2} - t_{1})
\end{equation}

Let us consider in particular the case $  Identification(
\mathbb{R} ; 0 , 2 \pi ) $ that is nothing but the circle $ S^{1}
$ endowed with the induced quotient topology.

Given $ f \in C^{\infty} ( \mathbb{R})$, $ t_{in} \in
\frac{\mathbb{R}}{\sim_{0,2 \pi }} $ and $ x_{in} \in \mathbb{R} $
as usual we will say that the Cauchy-problem:
\begin{equation}\label{eq:differential equation on circle}
  \dot{x} = f(x)
\end{equation}
\begin{equation} \label{eq:initial condition on the circle}
    x( t_{in} ) = x_{in}
\end{equation}
is well defined if and only it has one and only one solution $
\tilde{\tilde{x}} (t) : Identification( \mathbb{R} ; 0 , 2 \pi )
\mapsto ( \mathbb{R} , + , \cdot )$.

Denoted by $ \tilde{x} ( t)  : \mathbb{R} \mapsto \mathbb{R} $ the
solution of eq.\ref{eq:differential equation on circle}, eq.
\ref{eq:initial condition on the circle} for ordinary real time we
have clearly that:

\begin{proposition}
\end{proposition}
\begin{center}
The Cauchy problem of eq. \ref{eq:differential equation on
circle}, eq. \ref{eq:initial condition on the circle} is
well-defined if and only if:
\begin{equation}
   ( t_{1} \sim_{0 , 2 \pi} t_{2} \; \Rightarrow \;  \tilde{x} (t_{1}) = \tilde{x}
 (t_{2}) ) \; \; \forall t_{1},t_{2} \in \mathbb{R}
\end{equation}
 in which case the solution is:
\begin{equation} \label{eq:solution on the circle}
  \tilde{\tilde{x}} ( [ t ]_{\sim_{0 , 2 \pi}} ) \; := \; \tilde{x} (t) \; \; \forall
  t \in \mathbb{R}
\end{equation}
\end{center}
\begin{proof}
 The thesis immediately follows by the  definition \ref{def:identification of two points on the line}
\end{proof}

\bigskip

Let us then consider the introduction of causal circularities in
the topological space $ ( structure( 0 , 2 , 0) , \mathcal{T} ( 0
, 2 , 0) ) $;

at this purpose let us identify the point $ - \pi \in ( - \infty
,0) $ and the point $ + \pi_{(1)} \in [ 0_{(1)} , + \infty_{(1)} )
$ through the following:
\begin{definition} \label{def:identification of two point in structure 0 , 2 , 0}
\end{definition}
\emph{identification of $ - \pi $ and $  + \pi_{(1)} $ on $
structure( 0 , 2 , 0) $:}
\begin{equation}
Identification[  structure( 0 , 2 , 0) ; - \pi  , + \pi_{(1)} ] \;
:= \; \frac{(  structure( 0 , 2 , 0) ,
 \mathcal{T}( 0 , 2 ,0 ))}{ \sim_{- \pi , + \pi_{(1)} }}
\end{equation}
where $ \sim_{- \pi , + \pi_{(1)}} $ is the equivalence relation
over $ structure( 0 , 2 , 0)$ :
\begin{equation}
    a \sim_{- \pi , + \pi_{(1)}} b \; := \; \exists n \in \mathbb{Z} :
    b = a + n  [ \pi_{(1)} - ( - \pi ) ]
\end{equation}

\begin{definition}
\end{definition}
\emph{graph of $ Identification[  structure( 0 , 2 , 0) ; - \pi ,
+ \pi_{(1)}  ] $:}
\begin{center}
 $ graph \{ Identification[  structure( 0 , 2 , 0) ; - \pi ,
+ \pi_{(1)} ] \} := $ the graph obtained from $ graph [ structure(
0 , 2 , 0)  ] $ by the identification of the edge containing $  -
\pi $ and the edge containing $ + \pi_{(1)} $.
\end{center}

\smallskip

Given $ f \in C^{\infty} (  \mathbb{R}) $, $ t_{in} \in
\frac{structure( 0 , 2 , 0)}{\sim_{- \pi , + \pi_{(1)} }} $ and $
x_{in} \in \mathbb{R} $ as usual we will say that the
Cauchy-problem:
\begin{equation}\label{eq:differential equation on structure0,2,0 with causal circularity}
  \dot{x} = f(x)
\end{equation}
\begin{equation} \label{eq:initial condition on structure0,2,0 with causal circularity}
    x( t_{in} ) = x_{in}
\end{equation}
is well defined if and only it has one and only one solution $
\tilde{\tilde{x}} (t) : Identification[  structure( 0 , 2 , 0) ; -
\pi , + \pi_{(1)} ] \mapsto ( \mathbb{R} , + , \cdot )$.

Denoted by $ \tilde{x} ( t)  : \mathbb{R} \mapsto \mathbb{R} $ the
solution of eq. \ref{eq:differential equation on structure0,2,0
with causal circularity}, eq. \ref{eq:initial condition on the
circle} for ordinary real time we have clearly that:

\begin{proposition}
\end{proposition}
The Cauchy problem of eq. \ref{eq:differential equation on
structure0,2,0 with causal circularity}, eq. \ref{eq:initial
condition on structure0,2,0 with causal circularity} is
well-defined if and only if:
\begin{equation}
    a \sim_{- \pi , \pi_{(1)}} b \; \Rightarrow \;  \tilde{x} (a) = \tilde{x}
 (b )
\end{equation}
in which case the solution is:
\begin{equation} \label{eq:solution on structure0,2,0 with causal circularity}
  \tilde{\tilde{x}} ( [ t ]_{\sim_{- \pi , + \pi_{(1)}}} ) \; := \; \tilde{x} (t) \; \; \forall
  t \in \mathbb{R}
\end{equation}
\begin{proof}
 The thesis immediately follows by the definition \ref{def:identification of two point in structure 0 , 2 , 0}
\end{proof}

\bigskip

 Let us now consider a chronologically preordered structure $ structure(t_{1} , b_{1} , k_{1} ; t_{2,(i_{1})} ,
b_{2} , k_{2} ;  \cdots ; t_{n,(i_{n-1})} , b_{n} , k_{n} ) $ with
$ k_{1} := 1 $ and  $ k_{n} := 0 $ so that the graph $ graph\{
structure(t_{1} , b_{1} , k_{1} ; t_{2,(i_{1})} , b_{2} , k_{2} ;
\cdots ; t_{n,(i_{n-1})} , b_{n} , k_{n} ) \} $ has $ b_{1} $
input edges and $ b_{n} $ output edges.

Given $ t_{A,(i)} \in ( - \infty_{(i)} , t_{1,(i)} ) , i \in
\mathbb{N}_{+} : i \leq b_{1} $ and $ t_{B,(j)} \in ( t_{n,(j)} , +
\infty_{(j)} ) , j \in \mathbb{N}_{+} : j \leq b_{1} $:

\begin{definition} \label{def:identification of two external points on a chronologically preordered structure}
\end{definition}
\emph{identification of $ t_{A,(i)} $ and $ t_{B,(j)} $ on $
structure(t_{1} , b_{1} , k_{1} ; t_{2,(i_{1})} , b_{2} , k_{2} ;
\cdots ; t_{n,(i_{n-1})} , b_{n} , k_{n} ) $:}
\begin{multline}
Identification[   structure(t_{1} , b_{1} , k_{1} ; t_{2,(i_{1})}
, b_{2} , k_{2} ;  \cdots ; t_{n,(i_{n-1})} , b_{n} , k_{n} ) ;
t_{A,(i)} ,  t_{B,(j)} ] \; := \\
 \frac{(  structure(t_{1} , b_{1}
, k_{1} ; t_{2,(i_{1})} , b_{2} , k_{2} ;  \cdots ;
t_{n,(i_{n-1})} , b_{n} , k_{n} ) , \mathcal{T}(t_{1} , b_{1} ,
k_{1} ; t_{2,(i_{1})} , b_{2} , k_{2} ;  \cdots ; t_{n,(i_{n-1})}
, b_{n} , k_{n} ))}{ \sim_{ t_{A,(i)} ,  t_{B,(j)}  }}
\end{multline}
where $  \sim_{ t_{A,(i)} ,  t_{B,(j)}  } $ is the equivalence
relation over $  structure(t_{1} , b_{1} , k_{1} ; t_{2,(i_{1})} ,
b_{2} , k_{2} ;  \cdots ; t_{n,(i_{n-1})} , b_{n} , k_{n} )$ :
\begin{equation}
    a  \sim_{ t_{A,(i)} ,  t_{B,(j)}  }  b \; := \; \exists n \in \mathbb{Z} :
    b = a + n ( t_{B,(j)} - t_{A,(i)} )
\end{equation}

Given $ f \in C^{\infty} ( \mathbb{R} ) $, $ t_{in} \in
\frac{structure(t_{1} , b_{1} , k_{1} ; t_{2,(i_{1})} , b_{2} ,
k_{2} ;  \cdots ; t_{n,(i_{n-1})} , b_{n} , k_{n} )}{\sim_{
t_{A,(i)} ,  t_{B,(j)}}} $ and $ x_{in} \in \mathbb{R} $ as usual
we will say that the Cauchy-problem:
\begin{equation}\label{eq:differential equation on a  chronologically preordered structure with causal circularity}
  \dot{x} = f(x)
\end{equation}
\begin{equation} \label{eq:initial condition on a chronologically preordered structure with causal circularity}
    x( t_{in} ) = x_{in}
\end{equation}
is well defined if and only it has one and only one solution $
\tilde{\tilde{x}} (t) : Identification[   structure(t_{1} , b_{1}
, k_{1} ; t_{2,(i_{1})} , b_{2} , k_{2} ;  \cdots ;
t_{n,(i_{n-1})} , b_{n} , k_{n} ) ; t_{A,(i)} ,  t_{B,(j)} ]
\mapsto ( \mathbb{R} , + , \cdot )$.

Denoted by $ \tilde{x} ( t)  : \mathbb{R} \mapsto \mathbb{R} $ the
solution of eq. \ref{eq:differential equation on a chronologically
preordered structure with causal circularity}, eq. \ref{eq:initial
condition on a chronologically preordered structure with causal
circularity} for ordinary real time we have clearly that:

\begin{proposition}
\end{proposition}
The Cauchy problem of eq. \ref{eq:differential equation on a
chronologically preordered structure with causal circularity}, eq.
\ref{eq:initial condition on a chronologically preordered
structure with causal circularity} is well-defined if and only if:
\begin{equation}
    a \sim_{ t_{A,(i)} ,  t_{B,(j)}}  b \; \Rightarrow \;  \tilde{x} (a) = \tilde{x}
 (b )
\end{equation}
in which case the solution is:
\begin{equation} \label{eq:solution on a chronologically preordered structure with causal circularity}
  \tilde{\tilde{x}} ( [ t ]_{\sim_{ t_{A,(i)} ,  t_{B,(j)}}} ) \; := \; \tilde{x} (t) \; \; \forall
  t \in \mathbb{R}
\end{equation}
\begin{proof}
 The thesis immediately follows by the definition \ref{def:identification of two external points on a chronologically preordered structure}
\end{proof}

\smallskip

We can then suppose to perform more than an identification,
resulting in the following:
\begin{definition} \label{def:temporal structure}
\end{definition}
\emph{temporal structure:}
\begin{center}
  any topological space that can be obtained applying to a
  chronologically preordered temporal structure a finite number of
  identifications
\end{center}

\newpage
\section{Conclusions learnt from the toy model} \label{sec:Conclusions learnt from the toy model}
The analysis of the toy model performed in the previous sections
taught us that:
\begin{itemize}
    \item  as soon as the topological structure of time is
    chronologically preordered (so that the quotient with respect to such a preordering is a  partial, and in the occurring case actually linear,
    ordering):
    \begin{itemize}
        \item any single initial condition for the system represented by the model is
    the cause of its state at chronologically later times and is
    the effect of its state at chronologically previous times;
    parallel timelines are physically identical.
        \item multiple initial conditions at any time-division
        point are the cause of the state at chronologically later
        times and are the effect of the state at chronologically previous times
        if and only if they are identical, once again leading to physically identical parallel
        timelines; if such a consistency condition is not
        satisfied no causation occurs.
    \end{itemize}
    \item if chronological circularities are introduced in the
    topological structure of time, causation occurs if and only if
    suitable consistency conditions hold; in this case no
    mathematical contradiction arises from the existence of
    chronology violation \footnote{It should be remarked that in the toy model the existence of chronological violation may be defined as the condition
    that $ \preceq_{chronological} $ fails to be a preordering; hence, with this respect, the nonrelativistic toy model differs from the general relativistic situation in which, as stated by the proposition \ref{pr:properties of the chronological relation} , the chronological relation
    is still a preordering over the chronology violating set failing to be a partial ordering.}: in such a situation an event
    A may be both the cause and the effect of an event B without
    logical inconsistencies.
\end{itemize}
\newpage
\section{Visser's multihistory approach to time-travel paradoxes}

The first mathematical result interrelating the removal of the
assumption that space-time, as a topological space, has to be
Hausdorff and time-travel has been given (implicitly) by Petr
Hajicek in the Theorem 4 of \cite{Hajicek-71} that implies the
existence,  on a non Hausdorff space time $ ( M , g_{ab} ) $, of a
deep interrelation between the presence of chronology violations,
i.e. the situation in  which $ V_{chronology} ( M , g_{ab} ) \neq
\emptyset $, and the presence of time-like bifurcating paths.

Of particular interest is Hajicek's analysis about Cauchy problems
on non-Hausdorff topological spaces and bifurcating solutions that
we will analyze, again, in a simple model.

Given a non-Hausdorff topological space $ ( S , T ) $ let us
consider an alternative toy model, someway dual to the one
discussed in the previous sections (and that we will call
therefore the dual toy model), in that it is its configuration
space to be an non Hausdorff topological space, being $ (S ,T)$,
while time is assumed to be described by the topologically trivial
real line $ \mathbb{R} $. Let us suppose that such a dynamical
system is ruled by the differential equation $ \dot{x} = f(x) $
where f is a continuous function on $ (S,T)$.

Given $ t_{in} \in \mathbb{R} $ and $ x_{in} \in S $ Hajicek
analyzes the situation in which the Cauchy problem:
\begin{equation}
    \dot{x} \; = \; f(x)
\end{equation}
\begin{equation}
    x ( t_{in} ) \; = \; x_{in}
\end{equation}
has two solutions  $ \alpha_{1} $ and $ \alpha_{2} $ bifurcating
at a time $ t_{bif} \in \mathbb{R} \; : \; t_{bif} > t_{in} $ (see
appendix \ref{sec:Non Hausdorff topological spaces}).

According to our terminology such a dynamical system is still
deterministic though no causal relationships exist between its
states at different times.

\bigskip

Giving up  the assumption that space-time has to be Hausdorff
doesn't allow, by itself , to furnish a mathematical solution to
the so-called time-travel paradoxes belonging to Matt Visser's
option "Radical Rewrite Conjecture" \cite{Visser-03}, i.e.
allowing changes of the past by a free-will's owner.

An example of how this can be performed adding some other
ingredient to the permission of non-Hausdorff spacetimes has been
given by Matt Visser:

given a solution $ ( M , g_{ab} ) $  of Einstein's equation  such
that $ V_{chronoloy} ( M , g_{ab} ) \neq \emptyset $ Matt Visser's
multihistory approach to the emerging so-called time travel
paradoxes may be formalized as the addiction to General Relativity
of the following:

\begin{axiom} \label{ax:multihistory axiom in General Relativity}
\end{axiom}
 If a free-will's owner changes the past in $ p \in
V_{chronoloy} ( M , g_{ab} ) $ then the following phase transition
of  the Universe occurs:
\begin{equation}
    M  \; \rightarrow \;  Split ( M , \overline{J^{+}(p)} ,
    2)
\end{equation}

The analogous of axiom \ref{ax:multihistory axiom in General
Relativity} for our non relativistic toy model would consists in
the following:
\begin{axiom}
\end{axiom}
If a free-will's owner changes the past at time $ t_{1} \in X $
then the following phase transition in the topological structure
of time occurs:
\begin{equation}
    X \; \rightarrow \; Split ( X , t_{1} , 2)
\end{equation}
where X is the temporal structure describing time before the
action of the free-will's owner.

\smallskip

Let us call $ x_{1} $ the value that $ x( t_{1} ) $ would have had
if the free-will's owner hadn't changed the past at $ t_{1} $.

So the following generalized Cauchy problem faces us:
\begin{equation}
    \dot{x} \; = \; f(x)
\end{equation}
\begin{equation}
    x( t_{1,(1)} ) \; = \; x_{1}
\end{equation}
\begin{equation}
    x( t_{1,(2)}) \; = \; x_{2}
\end{equation}
where $ x_{2} \neq x_{1} $ is the new initial condition chosen by
the free-will's owner.

By the conclusions summarized in the section \ref{sec:Conclusions
learnt from the toy model} it follows that the changes of the past
break any causational relation.
\newpage
\section{Warning}
A slightly extended version of this paper (containing figures and
some example) is available at the author's homepage
http://www.gavrielsegre.com
\newpage
\appendix
\section{Preorderings} \label{sec:preorderings}
In this appendix we will briefly recall the basic notions
concerning relations over sets.

For the proof of the propositions, rather elementary, we demand to
\cite{Schechter-98}.

Given a set S:
\begin{definition}
\end{definition}
\emph{relation over S}:
\begin{equation}
    R \in \mathcal{P} ( S \times S )
\end{equation}
where $  \mathcal{P} ( S ) := \{ X : X \subseteq S \} $ is the
power-set of S.

Given a relation R over S let us introduce the useful notation:
\begin{equation}
    x R y \; := \; ( x , y) \in R
\end{equation}

Let us then introduce the following basic:
\begin{definition}
\end{definition}
\emph{R is a preordering:}
\begin{itemize}
    \item
\begin{equation}
    x R x \; \; \forall x \in S
\end{equation}
    \item
\begin{equation}
    ( x_{1} R x_{2} \: \wedge \: x_{2} R x_{3} \; \Rightarrow \; x_{1} R
    x_{3} ) \; \; \forall x_{1}, x_{2}, x_{3} \in S
\end{equation}
\end{itemize}

\begin{definition}
\end{definition}
\emph{equivalence relation over S:}
\begin{center}
 a preordering $ \sim $ over S such that:
 \begin{equation}
    (x_{1} \sim x_{2} \: \Rightarrow \: x_{2} \sim x_{1} ) \; \; \forall x_{1}, x_{2} \in S
\end{equation}
\end{center}
Given an equivalence relation $ \sim $ over S and $ x \in S$:
\begin{definition}
\end{definition}
\emph{equivalence class of x w.r.t. $ \sim $:}
\begin{equation}
    [x]_{\sim} \; := \; \{ y \in S \, : \, x \sim y \}
\end{equation}
\begin{definition}
\end{definition}
\emph{quotient of S w.r.t. $ \sim $:}
\begin{equation}
    \frac{S}{\sim} \; := \; \{ [ x ]_{\sim}  , x \in S \}
\end{equation}

\smallskip

Given a preordering $ \preceq $ over S and $ x_{1}, x_{2} \in S $:
\begin{definition}
\end{definition}
\begin{equation}
    x_{1} \sim_{\preceq} x_{2} \; := \; x_{1} \preceq x_{2} \;
    \wedge \; x_{2} \preceq x_{1}
\end{equation}

\begin{proposition}
\end{proposition}
\begin{center}
  $ \sim_{\preceq} $ is an equivalence relation over S
\end{center}

\smallskip

\begin{definition}
\end{definition}
\emph{partial ordering over S:}
\begin{center}
 a preordering $ \preceq $ over S such that:
\begin{equation}
    (x_{1} \sim_{\preceq} x_{2} \: \Rightarrow \: x_{1} = x_{2})
    \; \; \forall x_{1}, x_{2} \in S
\end{equation}
\end{center}

\begin{definition}
\end{definition}
\emph{linear ordering over S:}
\begin{center}
 a partial ordering $ \preceq $ over S such that:
\begin{equation}
    ( x_{1} \preceq x_{2} \: \vee \: x_{2} \preceq x_{1} )  \; \; \forall x_{1}, x_{2} \in S
\end{equation}
\end{center}

\newpage
\section{Non Hausdorff topological spaces} \label{sec:Non Hausdorff topological spaces}

In this appendix we will briefly recall the basic notions
concerning topological  spaces with a particular emphasis to non
Hausdorff ones.

For the proof of the propositions, rather elementary, we demand to
\cite{Geroch-85}.

 Given a set S:

\begin{definition}
\end{definition}
\emph{topology over S:}

$T \in \mathcal{P} ( S ) $ :
\begin{itemize}
    \item
\begin{equation}
    \emptyset , S \in T
\end{equation}
    \item
\begin{equation}
    O_{1} , O_{2} \in T \; \Rightarrow \;  O_{1} \cap O_{2} \in T
\end{equation}
    \item
\begin{equation}
    O_{i} \in  T  \,  \forall i \in I \; \Rightarrow \; \cup_{i \in I}
    O_{i} \in T
\end{equation}
\end{itemize}

where $  \mathcal{P} ( S ) := \{ X : X \subseteq S \} $ is the
power-set of S and where I is an arbitrary index set of arbitrary
cardinality.

We will denote the set of all the topologies over S by TOP(S).

Let us observe first of all that:
\begin{proposition}
\end{proposition}
\begin{itemize}
    \item discrete topology over S :=
\begin{equation}
  T_{discrete} (S) := \mathcal{P} ( S ) \in TOP(S)
\end{equation}
    \item indiscrete topology over S :=
\begin{equation}
 T_{indiscrete} (S) :=  \{ \emptyset , S \} \in TOP(S)
\end{equation}
\end{itemize}

Given $ T_{1} , T_{2} \in TOP (S)$:
\begin{definition}
\end{definition}
\emph{$ T_{1}$ is coarser than $ T_{2} $ ($ T_{2}$ is finer than $
T_{1} $) }
\begin{equation}
  T_{1} \preceq T_{2} \; := \;  T_{1} \subseteq T_{2}
\end{equation}

Given $ T \in TOP(S) $ and $ x_{1}, x_{2} \in S $
\begin{definition}
\end{definition}
\begin{equation}
    x_{1} \curlyvee x_{2} \; := \; O_{1} \cap  O_{2} \neq \emptyset \; \;  \forall O_{1},O_{2} \in T \; : \;
  x_{1} \in O_{1} \: \wedge \: x_{2} \in O_{2}
\end{equation}

\begin{definition}
\end{definition}
\emph{T is Hausdorff:}
\begin{equation}
  \neg ( x_{1} \curlyvee x_{2} ) \; \; \forall x_{1}, x_{2} \in S
\end{equation}

We will denote the set of all the Hausdorff topologies over S by $
TOP_{Hausdorff}(S) $.

\begin{proposition}
\end{proposition}
\begin{itemize}
    \item $ \preceq $ is a partial ordering over TOP(S)
    \item
\begin{equation}
   T_{indiscrete} (S) \preceq T \; \; \forall T \in TOP(S)
\end{equation}
    \item
\begin{equation}
    T \preceq T_{discrete} (S) \; \; \forall T \in TOP(S)
\end{equation}
 \item
 \begin{equation}
    T_{indiscrete} (S) \notin TOP_{Hausdorff}(S)
\end{equation}
 \item
 \begin{equation}
    T_{discrete} (S) \in TOP_{Hausdorff}(S)
\end{equation}
 \item
 \begin{equation}
    T_{1} \in TOP_{Hausdorff}(S) \, \wedge T_{1} \preceq T_{2} \;
    \Rightarrow \;  T_{2} \in TOP_{Hausdorff}(S)
\end{equation}
\end{itemize}

\begin{definition}
\end{definition}
\emph{topological space:}

a couple $ ( S , T ) $:
\begin{itemize}
    \item S is a set
    \item $ T \in TOP(S) $
\end{itemize}

Given a topological space $ ( S , T ) $ the elements of T are
called the open sets of $ ( S,T) $.

\begin{definition}
\end{definition}
\emph{$ ( S , T) $ is Hausdorff:}
\begin{equation}
    T \in  TOP_{Hausdorff}(S)
\end{equation}

Given $ A \subset S $:
\begin{definition}
\end{definition}
\emph{A is closed:}
\begin{equation}
    S - A \in T
\end{equation}
\begin{definition}
\end{definition}
\emph{closure of A:}
\begin{center}
  $ \bar{A} := $ the smallest closed set containing A
\end{center}
\begin{definition}
\end{definition}
\emph{interior of A:}
\begin{center}
 $  A^{\circ} := $ the largest open set contained in A
\end{center}
\begin{definition}
\end{definition}
\emph{boundary of A:}
\begin{equation}
    \partial A \; := \; \bar{A} -  A^{\circ}
\end{equation}

Given $  \{ O_{i} \}_{i \in I} \,  : \,   O_{i} \in T \,  \forall
i \in I $:

\begin{definition}
\end{definition}
\emph{ $  \{ O_{i} \}_{i \in I} $ is a base of $ ( S , T)$: }
\begin{equation}
    \forall O \in T \;  \exists I' : O = \cup_{i \in I'} O_{i}
\end{equation}

Clearly a base in a topological space individuates univocally the
underlying topology.

\begin{definition}
\end{definition}
\emph{natural topology over $ \mathbb{R} $:}
\begin{equation}
    T_{natural}( \mathbb{R} )  \in TOP( \mathbb{R} ) \; : \{  ( a , b) , a,b \in
    \mathbb{R} : a < b \} \text{ is a base of $ ( \mathbb{R} ,  T_{natural}( \mathbb{R} )) $}
\end{equation}

\begin{proposition}
\end{proposition}
\begin{equation}
   T_{natural}( \mathbb{R} )  \in  TOP_{Hausdorff}(S)
\end{equation}

We will assume from here and beyond that $ \mathbb{R} $, as a
topological space, is endowed with the natural topology
 $ T_{natural}( \mathbb{R} ) $.

\bigskip

Given a topological space $ X = ( S , T ) $ let us suppose to have
an equivalence relation $ \sim $ over S.
\begin{definition}
\end{definition}
\emph{quotient of X w.r.t. $ \sim $:}

\begin{center}
  the topological space $ \frac{X}{\sim} := (  \frac{S}{\sim} ,
  \frac{T}{\sim}) $ where $  \frac{T}{\sim} := \{ [ O ] : O \in T
  \} $
\end{center}

\begin{definition}
\end{definition}
\emph{The equivalence relation $ \sim $ is Hausdorff:}
\begin{center}
  $ \frac{X}{\sim} $ is an Hausdorff topological space
\end{center}

Given two equivalence relations $ \sim_{1} $ and $ \sim_{2} $ over
S:

\begin{definition}
\end{definition}
\emph{intersection of  $ \sim_{1} $ and $ \sim_{2} $:}
\begin{center}
  the equivalence relation $  \sim_{1} \wedge \sim_{2} $ over S
  such that:
\begin{equation}
    x  \sim_{1} \wedge \sim_{2} y \; := \; x \sim_{1} y \wedge  x \sim_{2} y
\end{equation}
\end{center}

Then:
\begin{proposition}
\end{proposition}
\begin{equation}
  \sim_{1} \: Hausdorff \; \wedge \;  \sim_{2} \: Hausdorff  \;
  \Rightarrow \; \sim_{1} \wedge \sim_{2} Hausdorff
\end{equation}

\bigskip

Given two topological spaces $ ( S_{1} , T_{1} ) $ and $ ( S_{2} ,
T_{2} ) $ and a map $ f : S_{1} \mapsto S_{2} $:

\begin{definition}
\end{definition}
\emph{f is continuous:}
\begin{equation}
 f^{-1} ( O) \in T_{1} \; \; \forall O \in T_{2}
\end{equation}

Given a topological space $ ( S , T ) $:

\begin{definition}
\end{definition}
\emph{path in (S,T)}:
\begin{equation}
     \alpha  :  ( \mathbb{R} , T_{natural} ( \mathbb{R} ) )  \mapsto ( S , T)  \text{ continuous }
\end{equation}

Given two path $ \alpha_{1} $ and $ \alpha_{2} $ in  (S,T) and a
number $ t_{bif} \in \mathbb{R} $:

\begin{definition} \label{def:bifurcating paths}
\end{definition}
\emph{ $ \alpha_{1} $ and $ \alpha_{2} $ bifurcate at $ t_{bif}
$:}
\begin{equation}
    \alpha_{1} ( t) = \alpha_{2} (t) \; \forall t \in ( - \infty , t_{bif} )
\end{equation}
\begin{equation}
    \alpha_{1} ( [ t_{bif} , + \infty ) ) \cap \alpha_{2} ( [ t_{bif} , + \infty ] ) \; =
    \; \emptyset
\end{equation}

\begin{proposition}
\end{proposition}
\begin{center}
    $ \exists \,  \alpha_{1} $ and $ \alpha_{2} $ paths in (S,T) bifurcating at $ t_{bif} \in
    \mathbb{R} \; \Rightarrow \; $ (S,T) is not Hausdorff
\end{center}
\begin{proof}
By definition \ref{def:bifurcating paths} $ \alpha_{1} ( t_{bif} )
\curlyvee \alpha_{2} ( t_{bif} ) $.
\end{proof}
\newpage
\section{Differential Calculus on non Hausdorff topological spaces} \label{sec:Differential Calculus on non Hausdorff topological spaces}

Given a set S which is the minimal suppletive mathematical
structure $ \mathcal{T} $ through which we have to endow S in
order that on $ (S , \mathcal{T}) $ differential calculus may be
defined ?

To answer such a deep question goes far beyond the goals of this
paper.

For our purposes it will be enough to observe that as to the
definition of limits, a metric structure is a too strong
requirement since the induced topology is Hausdorff; contrary the
assignment of a topology $ \mathcal{T} $ (not necessarily
Hausdorff) is sufficient:

given a map $ f : S \mapsto S $ on the topological space $ ( S ,
\mathcal{T} ) $ and two point $ x_{1}, x_{2} \in S $:
\begin{definition}
\end{definition}
\emph{f tends to $ x_{2} $ as x tends to $ x_{1} $:}
\begin{equation}
    \lim_{x \rightarrow x_{1}} f(x) = x_{2} \; := \; \forall O_{2}
    \in \mathcal{T} : x_{2} \in O_{2} \; \exists O_{1} \in
    \mathcal{T} \; : \; x_{1} \in O_{1} \, \wedge \, ( f(x) \in
    O_{2} \forall x \in O_{1} )
\end{equation}

In a similar way, given two sets $ S _{1} $ and $ S_{2} $, a map $
f : S_{1} \mapsto S_{2} $ and two point $ x_{1}, \in S_{1} $ and $
x_{2}, \in S_{2} $  the assignment of a topology $ \mathcal{T}_{1}
$ on $ S_{1} $ and of a topology $ \mathcal{T}_{2} $ on $ S_{2} $
is sufficient to define limits:
\begin{definition}
\end{definition}
\emph{f tends to $ x_{2} $ as x tends to $ x_{1} $:}
\begin{equation}
    \lim_{x \rightarrow x_{1}} f(x) = x_{2} \; := \; \forall O_{2}
    \in \mathcal{T}_{2} : x_{2} \in O_{2} \; \exists O_{1} \in
    \mathcal{T}_{1} \; : \; x_{1} \in O_{1} \, \wedge \, ( f(x) \in
    O_{2} \forall x \in O_{1} )
\end{equation}

\smallskip

As to the definition of the derivative of f the assignment of two
topologies on domain and codomain is not sufficient.

It is in fact necessary to require that $ S_{2} $ has the
algebraic structure of being a field w.r.t. two internal binary
operations $ + $ and $ \cdot $.

One can then introduce the following:
\begin{definition} \label{def:derivative}
\end{definition}
\emph{derivative of f in $ x_{1}$:}
\begin{equation}
    f'( x_{1} ) \; := \; \lim_{x \rightarrow x_{1}} \frac{f(x)-f(x_{1})}{ x-x_{1} }
\end{equation}

Clearly the definition \ref{def:derivative} may be iterated in
order to define derivatives of  of any order.

\end{document}